\documentclass[twocolumn,prl,showpacs,preprintnumbers,
               superscriptaddress,floatfix]{revtex4}

\usepackage{amssymb}
\usepackage{graphicx}
\usepackage{dcolumn}
\usepackage{bm}
\usepackage{longtable}

\newcommand{\etal}{{\em{et al}}}

\newcommand{\dfra}{\displaystyle\frac}
\newcommand{\ds}{$D^*_{sJ}$(2317)$^+$ }

\begin{document}

\title{\ds: a $P$ state from the light cone harmonic oscillator model?}

\author{Shan-Gui Zhou}
\email{sgzhou@mpi-hd.mpg.de}
\affiliation{Max-Planck-Institut f\"ur Kernphysik,
             69029 Heidelberg, Germany}
\affiliation{School of Physics, Peking University,
             Beijing 100871, China}
\affiliation{Institute of Theoretical Physics, Chinese Academy of Sciences,
             Beijing 100080, China}
\affiliation{Center of Theoretical Nuclear Physics, National Laboratory of
             Heavy Ion Accelerator, Lanzhou 730000, China}
\author{Hans-Christian Pauli}
\email{pauli@mpi-hd.mpg.de}
\affiliation{Max-Planck-Institut f\"ur Kernphysik,
             69029 Heidelberg, Germany}

\date{May 9, 2003}

\begin{abstract}
We show that the mass of the recently found meson, \ds could be
reproduced by an effective light cone Hamiltonian model with a
harmonic oscillator potential as confinement --- the light cone
harmonic oscillator model.
\end{abstract}


\maketitle

A recent experiment by the BABAR collaboration observes a new
narrow $c\bar s$ state, \ds with the invariant mass $M = 2.32$
GeV/$c^2$~\cite{BABAR03}. The state has natural spin-parity and
the low mass suggests a $J^\pi = 0^+$ assignment. In the
convention of the quark model (QM), correspondingly, $L=1$, $S=1$
and $J=0$, i.e., $^{2S+1}L_J$ = $^3P_0$. However, as argued in
Ref.~\cite{BABAR03}, both the small mass and the small width are
in conflict with most model predictions available then
~\cite{GK91,GI85,DE01}. For example, the mass of this state is
typically predicted between 2.4 and 2.6 GeV/$c^2$ in
Refs.~\cite{GK91,GI85,DE01} (cf. Table~\ref{tab:comp}). Due to
these conflictions, it is suggested in Ref.~\cite{BABAR03} either
to modify those models or to attribute this state, e.g., as a four
quark state. But we show here that \ds is still consistent with
the normal picture. The mass of \ds could be reproduced by an
effective light cone Hamiltonian model with a harmonic oscillator
potential as confinement --- the light cone harmonic oscillator
(LCH) model without changing the parameters fixed previously.

The LCH model, a light-cone QCD-inspired model, with the mass
squared operator consisting of a harmonic oscillator potential as
confinement and a Dirac delta interaction, was proposed
recently~\cite{FPZ02A,FPZ02B}. This simple model presents an
universal and satisfactory description of both singlet and triplet
states of $S$-wave mesons and the corresponding radial
excitations.

For the present purpose, the model Hamiltonian reads,
\begin{eqnarray}
 M^2_{\rm ho} \varphi (\vec r) = M^2 \varphi (\vec r) .
 \label{mass3}
\end{eqnarray}
Compared to Refs.~\cite{FPZ02A,FPZ02B}, the Dirac-delta
interaction is omitted here because it only acts on $S$ states.
But we add a phenomenological spin-orbit term with the strength
$\kappa$ being a free parameter to be determined by the data (For
the results with a more reasonable spin-orbit term, see
Ref.~\cite{ZHOU03b}). The mass squared operator is then
\begin{eqnarray}
 M^2_{\rm ho}= 2 m_s \left( - \dfra{\overrightarrow{\nabla}^2}{2m_r}
                            + \dfra{1}{2} m_s
                            + v(r) + \kappa \vec L \cdot\vec S
                     \right)
 \ ,
 \label{mass4}
\end{eqnarray}
in units of $\hbar = c = 1$. $m_s$ and $m_r$ are the sum and the
reduced mass of the two quarks with constituent quark masses $m_1$
and $m_2$. The masses for the strange and charm quarks were
determined in Ref.~\cite{FPZ02A,FPZ02B} as 0.478 and 1.749 GeV,
respectively.

The harmonic oscillator potential was introduced in
Ref.~\cite{FPZ02A,FPZ02B} as
\begin{equation}
 v(r) = - c_0 + \frac{1}{2} c_2 r^2
 ,
\end{equation}
where $c_0$ = 0.807 GeV and $c_2$ = 0.0713 GeV$^3$ are two
universal parameters valid for all mesons. The mass squared was
given analytically,
\begin{equation}
 M^2_{nLJ}
 = 2m_s \left[
         N \sqrt{\frac{c_2}{m_r}}
         + \frac{1}{2} m_s - c_0 + E_{LS}
        \right] ,
 \label{eq:masssquare}
\end{equation}
where $N = ( 2n+L+3/2 ) $ with $n$ = 0, 1, 2, $\cdots$ the radial
quantum number. The term of the spin-orbit splitting $E_{LS}$ is
supposed to vanish for $S=0$. For triplet states ($S=1$),
\begin{equation}
 E_{LS} = \left\{
           \begin{array}{ll}
           +\kappa L,     & \text{for $J=L+1$,} \\
           -\kappa,       & \text{for $J=L$,} \\
           -\kappa (L+1), & \text{for $J=L-1$.} \\
           \end{array}
          \right.
 \label{eq:sosplitting}
\end{equation}

\begin{table}[tbp]
\caption{Masses of $c\bar s$ meson states with $L = 1$. The data
for $^3P_2$ and $^1P_1$ are taken from Ref.~\protect\cite{PDG02}.
Predictions from Ref.~\protect\cite{GK91,DE01} are also included
for comparison. For the labels of each state, $n$ is the radial
quantum number, $L$ the total orbital angular momentum, $S$ the
total spin and $J$ the total angular momentum. $j$ is the total
angular momentum of the strange quark which is conserved in the
limit of large charm quark mass.} \label{tab:comp}
\begin{ruledtabular}
\begin{tabular}{cccccc}
 $n^{2S+1}L_J$ & $n^jL_J$   & Data  & Ref.~\protect\cite{GK91} &
 Ref.~\protect\cite{DE01} & This work  \\
\hline
 $0^3P_2$ & $0^{3/2}P_2$ & 2.573 & 2.59 & 2.581 & 2.573 \\
 $0^1P_1$ & $0^{3/2}P_1$ & 2.536 & 2.55 & 2.535 & 2.494 \\
 $0^3P_1$ & $0^{1/2}P_1$ & ---   & 2.55 & 2.605 & 2.412 \\
 $0^3P_0$ & $0^{1/2}P_0$ & 2.32  & 2.48 & 2.487 & 2.328 \\
\end{tabular}
\end{ruledtabular}
\end{table}

The mass of the singlet state $D_s(^1P_1)$, which is independent
of $\kappa$, is calculated from Eq.~(\ref{eq:masssquare}) as 2.494
GeV. It deviates from the experimental value $M(D_{s1}(2536)^+)$ =
2.536 GeV~\cite{PDG02} by only $\sim$0.04 GeV. The agreement is
reasonable considering that the mixing between the singlet and the
triplet $J=1$ states is not included in this simple model. For the
same reason, we will use the theoretical mass of $D_s(^1P_1)$ and
the data of $^3P_2$, $D^*_{sJ}(2573)^+$ to determine the parameter
$\kappa$,
\begin{equation}
 \kappa = \dfra{1}{2m_s} \left[ M^2(^3P_2) - M^2(^1P_1) \right]
        = 0.0899\ \text{GeV} \ ,
\end{equation}
where $m_s$ is the summation of the strange and charm quark
masses. The masses of the other two triplet $P$ states, $^3P_1$
and $^3P_0$ are easily calculated from Eqs.~(\ref{eq:masssquare})
and (\ref{eq:sosplitting}) and given in Table~\ref{tab:comp} where
predictions from Refs.~\cite{GK91,DE01} are also included for
comparison. The present prediction for the mass of the lowest $P$
state is very close to the data. This implies that \ds might still
be a ``normal'' meson consisting of two constituent quarks.

From the LCH model, the four $P$ states are equal-spacing in mass
squared as seen from Eqs.~(\ref{eq:masssquare}) and
(\ref{eq:sosplitting}) and Table~\ref{tab:comp}. Apparently this
is not the case in the experiment. As mentioned before, one of the
reasons might be that there is an admixture between $^1P_1$ and
$^3P_1$ because only the total angular momentum $\vec J = \vec L +
\vec S$ should be conserved. This coupling is not included in the
simple LCH model.

There are many other phenomenological scenarios for the spin-orbit
splitting in hadron spectroscopy, see e.g. Ref.~\cite{FS82}.
Inclusion of those scenarios would improve the prediction power of
our model but would also bring out more complication. We therefore
leave that for a future task.

In summary, the $P$ states of the charmed strange meson are
investigated by using an effective light cone Hamiltonian model
with a harmonic oscillator potential. We conclude that \ds can
still be consistent with the normal picture of mesons. The
investigation of other $L \ne 0$ states in all $q\bar q$ sectors
is in progress.

\begin{acknowledgments}
We thank Harun Omer for drawing our attention to
Ref.~\cite{BABAR03}. S. G. Z. was partly supported by the Major
State Basic Research Development Program of China Under Contract
Number G2000077407.
\end{acknowledgments}

\end{document}